\documentclass[iop]{emulateapj}

\usepackage{amsmath, amsthm, amssymb, slashed}
\usepackage{graphicx}
\usepackage[toc]{appendix}

\newcommand{\p}{^{\prime}}
\newcommand{\A}{{\cal A}}
\newcommand{\dd}{{\rm d}}
\newcommand{\nb}{{\bar{n}}}

\newcommand{\mpr}{m_{\rm p}}
\newcommand{\Int}{\int\limits}

\begin{document}


\title{An improved analytical model of the local interstellar magnetic
  field: \\ The extension to compressibility}

\shortauthors{Kleimann et al.}


\author{Jens Kleimann\altaffilmark{1}}
\affil{Ruhr-Universit\"at Bochum, Fakult\"at f\"ur Physik und Astronomie,
  Institut f\"ur Theoretische Physik IV, Bochum, Germany}
\email{jk@tp4.rub.de}

\author{Christian R\"oken} 
\affil{Universit\"at Regensburg, Fakult\"at f\"ur Mathematik, 
  Regensburg, Germany}  
\email{christian.roeken@mathematik.uni-regensburg.de}

\author{Horst Fichtner\altaffilmark{1}}
\affil{Ruhr-Universit\"at Bochum, Fakult\"at f\"ur Physik und Astronomie,
  Institut f\"ur Theoretische Physik IV, Bochum, Germany}
\email{hf@tp4.rub.de}

\altaffiltext{1}{Ruhr Astroparticle and Plasma Physics Center,
  Ruhr-Universit\"at Bochum, Germany}

\begin{abstract}
  A previously published analytical magnetohydrodynamic model for the local
  interstellar magnetic field in the vicinity of the heliopause
  (R\"oken et al.\ 2015) is extended from incompressible to compressible,
  yet predominantly subsonic flow, considering both isothermal and adiabatic
  equations of state.
  Exact expressions and suitable approximations for the density and the flow
  velocity are derived and discussed.  In addition to the stationary
  induction equation, these expressions also satisfy the momentum balance
  equation along stream lines. The practical usefulness of the corresponding,
  still exact analytical magnetic field solution is assessed by comparing it
  quantitatively to results from a fully self-consistent magnetohydrodynamic
  simulation of the interstellar magnetic field draping around the heliopause.
\end{abstract}

\section{Introduction}

Recently, the problem of an exact analytical magnetohydrodynamical (MHD)
solution for an idealized structure of the local interstellar magnetic field
draping around the heliopause was solved by \citet{Roeken_EA:2015},
hereafter referred to as Paper~I. (See \citet{Isenberg_EA:2015} for an
approximate, singularity-hampered solution.) It was obtained under the
assumptions that (i) the field is frozen into an (ii) axisymmetric and
(iii) incompressible interstellar plasma flow.
The assumption of axisymmetry has been dropped in the second paper of this
series \citep{Kleimann_EA:2016} by employing so-called distortion flows that
allow for more realistic cross sections of the heliospheric tail flattened
by the interstellar magnetic field. In that paper, the assumption of
incompressibility was addressed briefly by pointing out that the use of a
solenoidal distortion flow does not imply any constraints on the
compressibility of the interstellar plasma flows. In the present, third paper
of the series, we generalize the solution presented in Paper~I to the case of
a compressible plasma flow.   

The interest in such solutions lies in the fact that, on the one hand, the
interstellar plasma is likely to be supersonic and super-fast-magnetosonic
\citep{Ben-Jaffel_EA:2013, Scherer_Fichtner:2014} and, thus, is compressed at
an interstellar bow shock. (For an analytical treatment of bow shocks in
other astrophysical scenarios, see the recent paper by
\citet{Keshet_Naor:2016}.) The resulting subsonic low-Mach number flow in the
outer heliosheath between the heliopause and the bow shock is limited to a
relatively narrow region bounded by the so-called sonic lines
\citep[e.g.,][]{Scherer_EA:2016}, at which the flow again becomes supersonic
toward the flanks of the heliosphere. The ensuing flow compressibility in
this region has been discussed in the context of the stability of the
heliopause by, e.g., \citet{Caillol_Ruderman:2007} and
\citet{Belov_Ruderman:2010}. On the other hand, even for the case that such
a bow shock would not exist \citep{McComas_EA:2012}, the plasma flow in the
region of the so-called bow wave \citep{Zank_EA:2013} would, at least partly,
be characterized by Mach numbers below but close to unity
\citep{Gayley_EA:1997}, and the usual incompressibility assumption for
subsonic flows with low Mach numbers ($M \la 0.3$) would not hold. 
So, in any case, there are reasons to drop the strict incompressibility
assumption for the local interstellar medium.

As is demonstrated here, physically meaningful solutions for compressible,
yet predominantly subsonic flows can be worked out analytically. These
density structures and their approximations are derived and discussed in
Section~\ref{sec:density}. In Section~\ref{sec:mag-field}, we explicitly
compute the resulting improved magnetic field frozen into such an
approximative flow by exploiting the exact result of Paper~I, taking into
account modifications induced by the non-constant density. The usefulness of
this improved analytical field solution is assessed in
Section~\ref{sec:numerics} by comparing it to a fully self-consistent
numerical MHD model. A summary and conclusions are given in
Section~\ref{sec:summary}.
~\\

\section{Density Structure}
\label{sec:density}

In this section, we determine the physical (number) density structure of the
compressible model and discuss suitable approximations. \\

\subsection{The Rankine-type Heliosphere Model}

Since its introduction as a simple model for the interstellar flow in and
around the heliopause by \citet{Parker:1961}, the incompressible Rankine
half-body flow continues to be popular in the heliophysics community
\citep{Yu:1974, Nerney_Suess:1995, Fahr_EA:2014, Fahr_EA:2016,
  Isenberg_EA:2015, Sylla_Fichtner:2015, Zirnstein_McComas:2015}.
At its core lies a point-like mass source of strength $(4\pi u_0) q$ located
at the origin, superimposed on an undisturbed flow
\mbox{${\bf u}_0 = - u_0 \, {\bf e}_z$} that is incident from the $+z$
direction. The Rankine velocity field thus derives from a flow potential
\begin{equation}
  \label{eq:Phi_helio}
  \Phi({\bf r}) := u_0 \left( \frac{q}{r}+z \right)
\end{equation}
as
\begin{equation} \label{eq:u-flow}
  {\bf u}_{\rm R}({\bf r}) := - \nabla \Phi({\bf r}) = u_0 \left[
    \frac{q \, \rho}{r^3} \, {\bf e}_{\rho}
    + \left( \frac{q \, z}{r^3} -1 \right)
    {\bf e}_z \right] ,
\end{equation}
where $(\rho, \varphi, z)$ are cylindrical coordinates, and $r:=\|{\bf r}\|$.
It can easily be shown that the heliopause, defined as the set of all flow
lines emanating from the stagnation point $(\rho,z)_{\rm sp}:= (0,\sqrt{q})$,
is given by the surface
\begin{equation}
  \label{eq:def_hp}
  H(\rho, z) := 2q-\rho^2 - z \sqrt{4q - \rho^2} = 0 \ .
\end{equation}

The problem of an interstellar magnetic field being passively advected in the
Rankine flow (\ref{eq:u-flow}) was recently considered in Paper~I, which
provided the derivation of the exact analytical magnetic field solution of
the steady-state induction equation
\begin{equation}
  \label{eq:induct}
  \nabla \times [{\bf u}_{\rm R} \times {\bf B}] = {\bf 0}
\end{equation}
of ideal MHD and the magnetic divergence constraint
\begin{equation}
  \label{eq:div_b0}
  \nabla \cdot {\bf B} = 0 \ ,
\end{equation}
subject to boundary conditions consisting of an arbitrarily inclined,
homogeneous magnetic field at upstream infinity. \\

\subsection{The Extension to Compressibility}

While this idealized magnetic field was shown to yield a reasonable
approximation to corresponding results from a self-consistent numerical
model, one weakness that was identified in the comparison is that the pile-up
of magnetic flux ahead of the heliopause is restricted to a relatively narrow
layer, whereas this layer appears much broader in the numerical simulations,
at least if parameters are chosen such that no bow shock forms.
The cause of this shortcoming can be traced back to the fact that the
underlying flow field (\ref{eq:u-flow}) is incompressible and stems from an
ad-hoc choice which, however reasonable, does not honor any conservation laws
except that for mass.

In the improved model, which we present in this work, we relax the condition
of incompressibility in favor of a physically more realistic description,
while at the same time retaining as much as possible of the original flow
structure. The key idea here is that while we continue to employ the flow
potential (\ref{eq:Phi_helio}), it is now re-interpreted as a potential for
the particle flux density \mbox{${\bf s}:= n \, {\bf u}$}, rather than for
the velocity ${\bf u}$, where the number density $n$ is no longer constant,
but may vary in space, such that Eq.~(\ref{eq:u-flow}) is replaced by
\begin{equation}
  {\bf s} = -\nabla \Phi({\bf r}) \ .
\end{equation}
Consequently, we observe that
\begin{enumerate}
\item the flow line structure, and in particular the shape of the heliopause
  as given by Eq.~(\ref{eq:def_hp}), remains unchanged,\item the solenoidality
  of $\nabla \Phi$ now implies $\nabla \cdot (n \, {\bf u}) = 0$, i.e., mass
  continues to be conserved, and
\item the flow ${\bf u}$ itself is no longer incompressible (or irrotational),
  and the density may vary along stream lines.
\end{enumerate}

Being derived from a global potential, the particle flux density is bound to
be differentiable in the entire domain of interest. This implies that our
model cannot accommodate oblique shocks, and thus, in particular, no bow
shock ahead of which ${\bf s}$ could be genuinely undisturbed (i.e.,
constant); it is a model for predominantly subsonic flow.
However, it is worth noting that the use of a flow potential as such does not
preclude supersonic or even transsonic flow as long as the physical quantity
that is computed from the potential's gradient can rightfully be taken as
irrotational and void of boundary layers
\citep[see, e.g.,][]{Curle_Davies:1971, Caughey:1982}.
Indeed, the model presented here does allow for smooth transitions to
moderately supersonic velocities in the heliotail's flanks, as will be shown
in Section~\ref{sec:approx}. 
Moreover, the absence of shocks should not be viewed as too severe a
restriction in its applicability to the real heliosphere, the outer flow of
which is clearly supersonic but may or may not be super-Alfv\'enic. The
existence of a bow shock depends on the flow being faster than the
fastest-propagating signal, which in this case are fast magnetosonic waves.
Therefore, the upstream Mach number $m$ that will be introduced later in this
section should not so much be identified with the actual sonic Mach number of
the incoming interstellar flow, but rather be viewed as a parameter to be
chosen close to the fast magnetosonic Mach number, which may well be around
or even below unity.

Since, in this new framework, we have so far only fixed the product of $n$
and ${\bf u}$, an additional equation is needed. One of the most
straightforward choices would the momentum balance equation
\begin{equation}
  \label{eq:momentum}
  \mpr n \, ({\bf u} \cdot \nabla) {\bf u} = - \nabla P
\end{equation}
with $\mpr$ being the proton mass and $P$ the gas pressure. (We refrain from
introducing the otherwise canonical symbol $\rho$ for the density $\mpr n$ in
order to reserve it for the cylindrical radius.) This yields two differential
equations for $n$, as the azimuthal component vanishes identically due to
rotational symmetry. For our choice of ${\bf s}$ there is no solution for $n$
satisfying both equations simultaneously. Rather than arbitrarily picking one
of them, we instead consider the projection of the momentum equation onto
stream lines.
This is essentially identical to the use of Bernoulli's equation for
compressible, adiabatic flow \citep[e.g.,][]{Fahr_Neutsch:1983}. The system
of equations is then to be closed with the polytropic equation of state
\begin{equation}
  \label{eq:adiabatic_eos}
  P(n) = C \, n^{\gamma} \ ,
\end{equation}
in which $\gamma$ is the adiabatic index, and the constant $C$ is determined
from a boundary condition at infinity, see Eq.~(\ref{eq:c_const}).

The projection of Eq.~(\ref{eq:momentum}) onto stream lines reads 
\begin{eqnarray} \label{eq:s2_long}
  - {\bf s} \cdot \nabla P &=& {\bf s} \cdot
  \left[ \mpr n \ ({\bf u} \cdot \nabla) {\bf u} \right] \nonumber \\
  &=& \mpr n \ {\bf s} \cdot
  \left[ \frac{1}{2} \nabla \left( {\bf u}^2 \right) -
    {\bf u} \times (\nabla \times {\bf u}) \right] \nonumber \\
  &=& \frac{\mpr n}{2} \ {\bf s} \cdot \nabla
  \left(\frac{{\bf s}^2}{n^2}\right)
  - \mpr \underbrace{{\bf s} \cdot \left[ {\bf s} \times
      \left( \nabla \times {\bf u} \right) \right]}_{=0} \nonumber \\
  &=& \frac{\mpr n}{2} \ {\bf s} \cdot
  \left[ \left( -\frac{2}{n^3} \nabla n \right) {\bf s}^2
    + \frac{1}{n^2} \nabla \left( {\bf s}^2 \right) \right]  \\
  &=& - \frac{\mpr}{n^2} \left( {\bf s} \cdot \nabla n \right) {\bf s}^2
  + \frac{\mpr}{2 \, n} \ {\bf s} \cdot \nabla \left( {\bf s}^2 \right) \ .
  \nonumber
\end{eqnarray}
Substituting the gradient of Eq.~(\ref{eq:adiabatic_eos})
\begin{equation}
  \nabla P = \left( \frac{\dd P}{\dd n} \right) \nabla n
  = C \gamma \, n^{\gamma-1} \nabla n
\end{equation}
into the left-hand side of Eq.~(\ref{eq:s2_long}) yields
\begin{equation}
  \label{eq:ODE}
  \quad 2 \left[ \frac{{\bf s}^2}{n} 
    - \frac{C \gamma}{\mpr} \ n^{\gamma} \right]
  {\bf s} \cdot \nabla n  = {\bf s} \cdot \nabla \left( {\bf s}^2 \right) \ .
\end{equation}
Since ${\bf s}$ is prescribed, Eq.~(\ref{eq:ODE}) constitutes a nonlinear
first-order partial differential equation for $n$ that will now be solved
using the method of characteristics. To this end, we introduce new
coordinates $\alpha$, $\beta$ such that
\begin{equation}
  \label{eq:s_d_nabla}
  {\bf s} \cdot \nabla = s_{\rho} \, \partial_{\rho} + s_z \, \partial_z
  = p \, \partial_{\alpha} = p \left( \frac{\dd \rho}{\dd \alpha} \,
    \partial_{\rho} + \frac{\dd z}{\dd \alpha} \, \partial_z \right)
\end{equation}
with an arbitrary function $p = p(\alpha, \beta)$. This condition results in
the coupled system of ordinary differential equations (ODEs)
\begin{equation}
  \frac{\dd \rho}{\dd \alpha} = \frac{s_{\rho}}{p}
  \quad \mbox{and} \quad\ 
  \frac{\dd z}{\dd \alpha} = \frac{s_z}{p} \ .
\end{equation}
Choosing $p = s_{\rho}$ yields
\begin{equation}
  \frac{\dd \rho}{\dd \alpha} = 1 \quad \Leftrightarrow \quad
  \rho = \alpha + \mathcal{F}(\beta)
\end{equation}
and in particular
\begin{equation}
  \label{eq:def_flowline}
  \frac{\dd z}{\dd \alpha} = \frac{s_z}{s_{\rho}} \ ,
\end{equation}
which implies that the characteristics and the stream lines of ${\bf s}$
coincide. We may thus replace the operator ${\bf s} \cdot \nabla$ in
Eq.~(\ref{eq:ODE}) by the derivative $s_{\rho} \, \partial_{\alpha}$ along the
stream lines. After multiplication with
$- \mpr/ \big( C \gamma n_0^{\gamma + 1} s_{\rho} \big)$, where $n_0$ is the
number density at upstream infinity, we find
\begin{equation}
  \label{eq:dn_dalpha}
  \left[ -\frac{\mpr}{C \gamma n_0^{\gamma+1}} \ {\bf s}^2
    + \left(\frac{n}{n_0}\right)^{\gamma+1} \right]
  \frac{2}{n} \, \frac{\partial n}{\partial \alpha}
  = -\frac{\mpr}{C \gamma n_0^{\gamma+1}} \,
  \frac{\partial {\bf s}^2}{\partial \alpha} \ .
\end{equation}
In terms of the function
\begin{equation}
  \label{eq:def_g}
  g := -\frac{\mpr}{C \gamma \, n_0^{\gamma+1}} \ {\bf s}^2 \ ,
\end{equation}
Eq.~(\ref{eq:dn_dalpha}) can be rewritten as
\begin{equation}
  \left[g + \left(\frac{n}{n_0}\right)^{\gamma+1}\right]
  \frac{2}{n} \, \frac{\partial n}{\partial \alpha} =
  \frac{\partial g}{\partial \alpha} \ .
\end{equation}
Using $g$ --- instead of $\alpha$ --- as a new coordinate, we apply the chain
rule to arrive at the ODE
\begin{equation}
  \label{eq:ode_nb}
  \left[ g + \nb^{\gamma + 1} \right] 
  \frac{2}{\nb} \, \frac{\dd \nb}{\dd g} = 1
\end{equation}
for the normalized density $\bar{n} := n/n_0$. (For the remainder of this
paper, a subscript $0$ marks quantities taken at upstream infinity, and a bar
denotes normalization with respect to this boundary value, i.e.,\
\mbox{$\bar{X} = X/X_0$} for any quantity $X$.) Introducing the sound speed
\begin{equation}
  \label{eq:def_cs}
  c := \left( \frac{\dd P}{\dd(\mpr n)} \right)^{1/2}
  = \left( \frac{C \gamma}{\mpr} n^{\gamma-1} \right)^{1/2} \ ,
\end{equation}
and evaluating this expression at upstream infinity, the denominator of $g$
in (\ref{eq:def_g}) becomes
\begin{equation}
  \label{eq:c_const}
  C \gamma \, n_0^{\gamma+1} = \mpr \, n_0^2 \, c_0^2 \ .
\end{equation}
Furthermore, since ${\bf s}_0^2 = (n_0 \, {\bf u}_0)^2$, we may write
(\ref{eq:def_g}) as
\begin{equation}
  \label{eq:n_a}
  g = - m^2 \, \frac{{\bf s}^2}{{\bf s}_0^2} = - m^2 \, \bar{\bf s}^2 \ , 
\end{equation}
in which the parameter $m := \|{\bf u}_0\|/c_0$ denotes the hydrodynamic Mach
number at upstream infinity, and the factor $\bar{\bf s}^2$ evaluates to
\begin{equation}
  \label{eq:defcalA}
  \begin{split}
    \bar{\bf s}^2 =&\ \frac{1}{{\bf s}_0^2}
    \left( -\nabla \Phi \right)^2
    = \left[ \left( \frac{q \rho}{r^3}    \right) {\bf e}_{\rho}
      +      \left( \frac{q z   }{r^3} -1 \right) {\bf e}_z \right]^2 \\
    =&\ 1 - \frac{2 q z}{r^3} + \frac{q^2}{r^4} =: \A \ .
  \end{split}
\end{equation}
This quantity attains its minimum value of zero at the stagnation point
$(\rho,z)=(0, \sqrt{q})$ and tends to unity both for $r \rightarrow \infty$
and on the surface $2 r z = q$, which passes through
$(\rho,z)=(0,\sqrt{q/2})$ and approaches the $z=0$ plane for large $r$.
The largest value outside the heliopause is reached on this surface at
$(\rho,z)=( 2\sqrt{6}/3 , \, -\sqrt{3}/3 ) \, \sqrt{q}$ and amounts to $4/3$.
In other words, $\A$ maps the entire region exterior to the heliopause onto
the interval $[0, 4/3]$. \\

\subsection{Analytical Solutions}

In order to solve Eq.~(\ref{eq:ode_nb}), we rewrite it as an ODE for the
function $g(\nb)$. To this end, we need to establish that $\nb(g)$ is
invertible.  Since, for $\nb$ being continuous, this is equivalent to
$(\dd \nb/\dd g)$ never changing sign, we see from Eq.~(\ref{eq:ode_nb}) that
the square bracket must not pass through zero. Since it evaluates to
\begin{equation}
  \begin{split}
    g + \nb^{\gamma+1} =&\ -m^2 \, \frac{{\bf s}^2}{{\bf s}_0^2}
    + \left(\frac{n}{n_0} \right)^{\gamma+1} \\
    =&\ - \frac{{\bf u}_0^2}{c_0^2} \,
    \frac{(n \, {\bf u})^2}{(n_0 \, {\bf u}_0)^2}
    + \left(\frac{n}{n_0}\right)^{2} \left(\frac{c}{c_0}\right)^{2} \\
    =&\ \left(\frac{n}{n_0}\right)^{2} \, \frac{c^2-{\bf u}^2}{c_0^2} \ ,
  \end{split}
\end{equation}
such a change of sign could only occur at a sonic transition. This implies
that our model's momentum-conserving flow solution is restricted
to the purely subsonic case
${\bf u}^2 < c^2 \ \Rightarrow \ m<1$.

Having established that the inverse $\nb^{-1} = g$ exists, we multiply
Eq.~(\ref{eq:ode_nb}) by $(\dd g/\dd \nb)$ and again apply the chain rule,
which yields
\begin{equation}
  \label{eq:ODE4}
  \frac{\dd g}{\dd \nb} - \frac{2 \, g}{\nb} = 2 \, \nb^{\gamma} \ .
\end{equation}
The homogeneous solution, i.e., the solution of the ODE
\begin{equation} 
  \frac{\dd  g_{\rm hom}}{\dd \nb} - \frac{2 \, g_{\rm hom}}{\nb} = 0 \ ,
\end{equation}
is simply obtained by integration with respect to $\nb$ as
\begin{equation} 
  g_{\rm hom} = c_1 \, \nb^2 \ ,
\end{equation}
where $c_1 \in \mathbb{R}$ is a constant. Accordingly, the inhomogeneous
solution can be derived using the ansatz $g_{\rm inhom} = h(\nb) \, \nb^2$,
leading to the ODE
\begin{equation} 
  \frac{\dd h}{\dd \nb} = 2 \, \nb^{\gamma - 2}
\end{equation}
for $h$, which can also be solved by simple integration with respect to $\nb$.
The inhomogeneous solution $g_{\rm inhom}$ is then
\begin{equation}
  \label{eq:inhomsol}
  g_{\rm inhom} = \nb^2 \times \left\{
    \begin{array}{ccc}
      \ln \left( \nb^2 \right) + c_2 &:& \gamma = 1 \\   && \\     
      \displaystyle  \frac{2 \, \nb^{\gamma-1}}{\gamma-1} + c_2 &:&
      \gamma \ne 1
    \end{array} \right.
\end{equation}
with a constant $c_2 \in \mathbb{R}$. Since the full solution is a
superposition of the homogeneous and the inhomogeneous solution, and the
constant $c_1$ can be absorbed into the constant $c_2$, it follows that $g$
itself is of the form (\ref{eq:inhomsol}).
Finally, the density can be determined by ``solving'' the transcendental
equations in (\ref{eq:inhomsol}) for $\nb$. This is done in the following
two subsections, considering separately the isothermal \mbox{($\gamma = 1$)}
and the adiabatic \mbox{($\gamma = 5/3$)} cases. \\

\subsection{Isothermal Flow}

Inserting the ansatz
\begin{equation}
  \nb(g) = \exp \left[ \frac{F \big(g \, \exp(c_2)  \big)}{2} \right]
\end{equation}
into Eq.~(\ref{eq:inhomsol}) for $\gamma = 1$ results in
\begin{equation}
  g = \left[ F \big(g \, \exp(c_2)  \big) + c_2 \right] \,
  \exp \left[ F \big(g \, \exp(c_2)  \big) \right] \ .
\end{equation}
Using the abbreviations
\begin{eqnarray}
  \tilde{g}    &:=& g \, \exp(c_2) \\
  W(\tilde{g}) &:=& F(\tilde{g}) + c_2 \ , 
\end{eqnarray}
this can also be written as
\begin{equation}
  \label{eq:def_W}
  \tilde{g} = W(\tilde{g}) \, \exp \big[ W(\tilde{g}) \big] \ ,
\end{equation}
which is the defining equation for Lambert's W function. Thus, we obtain
\begin{equation} \label{eq:n_iso}
  n = n_0 \exp \left( \frac{W \big[g \, \exp(c_2) \big] - c_2}{2} \right)
\end{equation}
for the density. The range of density values to be covered stipulates that
the principal branch of $W$ be employed.
Furthermore, the constant $c_2$ is fixed by the boundary values
\begin{equation}
  \label{eq:BC_c2}
  \begin{split}
    & \lim_{n \rightarrow n_0}{g}
    = \lim_{r \rightarrow \infty} (-m^2) \, \bar{\bf s}^2
    = - m^2 \\
    & \lim_{n \rightarrow n_0}{g}
    = \lim_{\nb \rightarrow 1} \nb^2 \left[ \ln{\left(\nb^2\right)} + c_2\right]
    = c_2
  \end{split}
\end{equation}
according to Eqs.~(\ref{eq:n_a}) and (\ref{eq:inhomsol}). It then directly
follows that $c_2 = - m^2$. In cylindrical coordinates, the density finally
becomes
\begin{equation}
  \label{eq:nA_isotherm}
  \begin{split}
    & n(\rho, z) = \\ & n_0 \exp \left[ \frac{m^2}{2} + \frac{1}{2}
      W \left( - \frac{m^2}{\exp(m^2)} \, \A(\rho,z)
      \right) \right]
  \end{split}
\end{equation}
with $\A$ defined in (\ref{eq:defcalA}). \\

\subsection{ Mono-atomic Ideal Gas Flow}

For $\gamma \in \mathbb{N} \backslash \{1\}$, Eq.~(\ref{eq:inhomsol}) gives
the zeros of a polynomial in $\nb$ with powers of 0, 2, and $\gamma+1$.
Analytical inversion is therefore only possible for $\gamma \in \{2,3 \}$,
and none of these cases is particularly meaningful in the heliospheric
context. Moreover, since we are mainly interested in the case of
$\gamma = 5/3$ describing mono-atomic ideal gases, we substitute this value
into the adiabatic flow equation (cf.\ the second case of
Eq.~(\ref{eq:inhomsol})), obtaining 
\begin{equation}
  g = \nb^2 \big( 3 \, \nb^{2/3} + c_2 \big) \ . 
\end{equation}
In analogy to (\ref{eq:BC_c2}), the constant $c_2$ is found from
\begin{equation}
  -m^2 = \lim_{\nb \rightarrow 1}{g} = 3 + c_2  \quad
  \Leftrightarrow \quad c_2 = - \big( m^2 + 3 \big) \ .
\end{equation}
The density is then implicitly given by 
\begin{equation}
  n(g) = n_0 \, \nb(g)
\end{equation}
with $\nb$ being the solution to 
\begin{equation}
  \label{eq:nA_non-isoth}
  g = \nb^2 \left[ 3 \big( \nb^{2/3} - 1 \big) - m^2 \right] \ .
\end{equation}
In a cylindrical representation, the density is again obtained by replacing
$g$ with $-m^2 \, \A$.

  \begin{center}
    \begin{figure*}[t]
      \includegraphics[width=\textwidth]{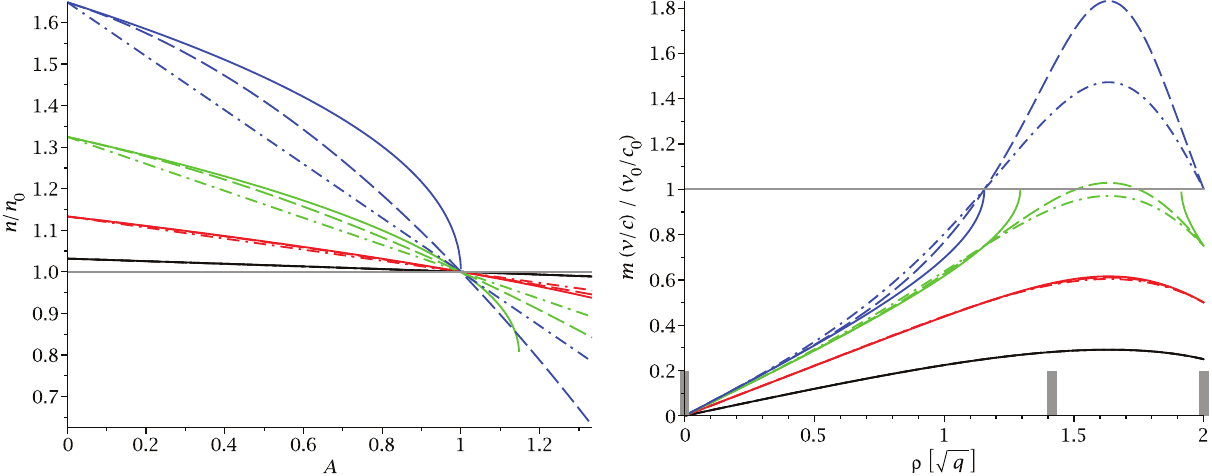}
      \caption{\label{fig:n-A_approx}
        Left: density profile (\ref{eq:imp_gamma_eq1}) (solid) vs.\ its
        approximation of first (dashed-dotted) and second (dashed) order
        according to Eqs.~(\ref{eq:n1_app}) and (\ref{eq:n2_app}) for
        $m=0.25$ (black), $m=0.5$ (red), $m=0.75$ (green), and $m=1.0$ (blue)
        in the domain of interest ${\cal A} \in [0,4/3]$. Note that in the
        $m=0.75$ case, the exact solution cannot be continued beyond
        ${\cal A}_{\rm max} \approx 1.021$ due to constraint (\ref{eq:Amax})
        (and for $m=1.0$ beyond ${\cal A}_{\rm max}=1$ for the same reason),
        whereas both approximations are well-defined on the entire domain.
        Right: local Mach number along the heliopause for the same cases (and
        using the same color mapping) as on the left, parameterized by $\rho$
        (i.e., with the $z$ coordinate chosen such that Eq.~(\ref{eq:def_hp})
        is satisfied). The thick, gray tickmarks at
        $\rho/\sqrt{q} \in \{ 0,\sqrt{2}, 2 \}$ indicate the stagnation
        point ($z=\sqrt{q}$), the crosswind direction ($z=0$), and downwind
        infinity ($z \rightarrow -\infty$), respectively. In both plots,
        additional horizontal lines at unity have been inserted to guide
        the eye. \\
      }
    \end{figure*}
  \end{center}

\subsection{Density Approximations}
\label{sec:approx}

The formulas for the density profiles
\begin{equation}
  \label{eq:imp_gamma_eq1}
  \nb(\A) = \exp \left[ \frac{m^2}{2}
    + \frac{1}{2} W \left(- \frac{m^2}{\exp(m^2)} \, \A
    \right) \right]
\end{equation}
for $\gamma = 1$ (cf.\ Eq.~(\ref{eq:nA_isotherm})) and
\begin{equation}
  \label{eq:imp_gamma_ne1}
  \nb^2 \left[ 1- \frac{3}{m^2} \big( \nb^{2/3}-1 \big) \right] = \A
\end{equation}
for $\gamma =5/3$ (cf.\ Eq.~(\ref{eq:nA_non-isoth})) are somewhat cumbersome
to handle due to the involvement of Lambert's W function, which is only
implicitly defined by the transcendental equation (\ref{eq:def_W}), and the
fact that Eq.~(\ref{eq:nA_non-isoth}) leads to a fourth-order polynomial in
$\nb^2$.  For this reason, we consider in the following suitable analytical
approximations for $\nb(\A)$ and the corresponding approximate flow fields.
These open up the possibility to find exact analytical magnetic field
solutions of the steady-state induction equation (\ref{eq:induct}) and the
magnetic divergence constraint (\ref{eq:div_b0}) with respect to these
approximative flows.

Both density profiles (\ref{eq:imp_gamma_eq1}) and (\ref{eq:imp_gamma_ne1})
can be reasonably well approximated by linear functions $\nb^{(1)}(\A)$ that
pass through $\nb(1)=1$ (thus ensuring $n \rightarrow n_0$ at infinity,
where we prescribe the boundary conditions) and the maximum value
$\nb(0) =: \nb_{\rm sp}$ reached at the stagnation point.
Although in principle any other values of $\A \in [0,4/3]$ could be used,
values close to $4/3$ are only approached for large, finite distances in
$(\rho,z)$ space and, therefore, of minor relevance for our intended
application. Using second-order polynomials $\nb^{(2)}(\A)$, the additional
degree of freedom may be fixed by requiring the correct derivative
$D_{\A_{\rm C}} := \partial_{\A} \nb|_{\A_{\rm C}}$ at either point
$\A_{\rm C} \in \{0,1 \}$ (or, alternatively, the correct density at some
intermediate value of $\A$).
We settle for a correct derivative at the stagnation point $\A=0$. In this
case, suitable approximations for $\nb(\A)$ to first and second order are
\begin{eqnarray}
  \label{eq:n1_app} \nb^{(1)}(\A) &:=&\
  \nb_{\rm sp} - (\nb_{\rm sp}-1) \, \A \\
  \label{eq:n2_app} \nb^{(2)}(\A) &:=&\ \nb_{\rm sp}+D_0 \, \A
  - ( \nb_{\rm sp}-1 +D_0) \, \A^2
\end{eqnarray}
with
\begin{equation}
  \label{eq:n_max}
  \nb_{\rm sp} = \left\{ \begin{array}{lcl}
      \exp \left(m^2/2 \right)     &\ : \ & \gamma=1 \\
      \left( 1+m^2/3 \right)^{3/2} &\ : \ & \gamma=5/3
    \end{array} \right. 
\end{equation}
and
\begin{equation}
  \label{eq:D_0}
  D_0 = -\frac{m^2}{2} \times
  \left\{ \begin{array}{lcl}
      \exp \left(-m^2/2 \right) &\ : \ & \gamma=1 \\  \displaystyle
      \left( 1+m^2/3 \right)^{-5/2}
      &\ : \ & \gamma=5/3 \ , \quad
    \end{array} \right.
\end{equation}
for which we have used the fact that Eq.~(\ref{eq:ode_nb}) at $\A=0$ can be
transformed into
\begin{equation}
  \left. \frac{\partial \nb}{\partial \A} \right|_{\A=0}
  = \frac{1}{2} \left.
    \left( \frac{\A}{\nb}-\frac{\nb^{\gamma}}{m^2}\right)^{-1} \right|_{\A=0}
  = -\frac{m^2}{2 \, (\nb_{\rm sp})^{\gamma}}
\end{equation}
and evaluated separately for both values of $\gamma$. For instance, when
choosing the specific Mach number \mbox{$m=0.6$}, we obtain
\begin{equation}
  \label{eq:n_app_vals}
  \nb^{(2)}(\A)|_{0.6} = \left\{ \begin{array}{lcl}
      1.20 -(\A/6.65) -(\A/4.62)^2 &:& \gamma=1 \\
      1.19 -(\A/7.38) -(\A/4.49)^2 &:& \gamma=5/3
    \end{array} \right.
\end{equation}
as suitable approximations for the respective density profiles.

As can be deduced from (\ref{eq:n_max}), the peak density $\nb_{\rm sp}$ is
always slightly higher in the isothermal case (by a factor of about
$(1+m^4/14)$, i.e., at most $\sim{}7$\%). This is plausible, since in this
case the pressure gradient that decelerates the incoming flow ahead of the
stagnation point is not caused by the combined gradients of density and
temperature, but has to come from density alone.

Fig.~\ref{fig:n-A_approx} compares the exact isothermal solution for various
values of $m$ against its first- and second-order approximations.
As expected, the agreement is most favorable for small $m$, but continues to
be useful also for larger values. The corresponding plot for $\gamma=5/3$ is
not shown since it would look very similar to its isothermal counterpart.

It should be noted that, since Lambert's W function can, by definition, only
accommodate real-valued arguments larger than or equal to $-1/e$
(corresponding to subsonic flow), the density (\ref{eq:imp_gamma_eq1}) ceases
to be well-defined in regions for which
\begin{equation}
  \label{eq:Amax}
  \A > {\cal A}_{\rm max}:= \exp(m^2-1)/m^2 \ . 
\end{equation}
However, since $\A \le 4/3$ holds everywhere outside the heliopause, such
regions can only arise for Mach numbers larger than
$m_{\rm crit, isoth} := \sqrt{-W(-3/ (4 \, e))} \approx 0.648$
(see, for instance, the solid curves for $m=0.75$ and $m=1.0$).
For \mbox{$\gamma=5/3$}, a corresponding critical Mach number is found at
$m_{\rm crit, adiab} \approx 0.619$.
\begin{figure}
  \includegraphics[width=8.5cm]{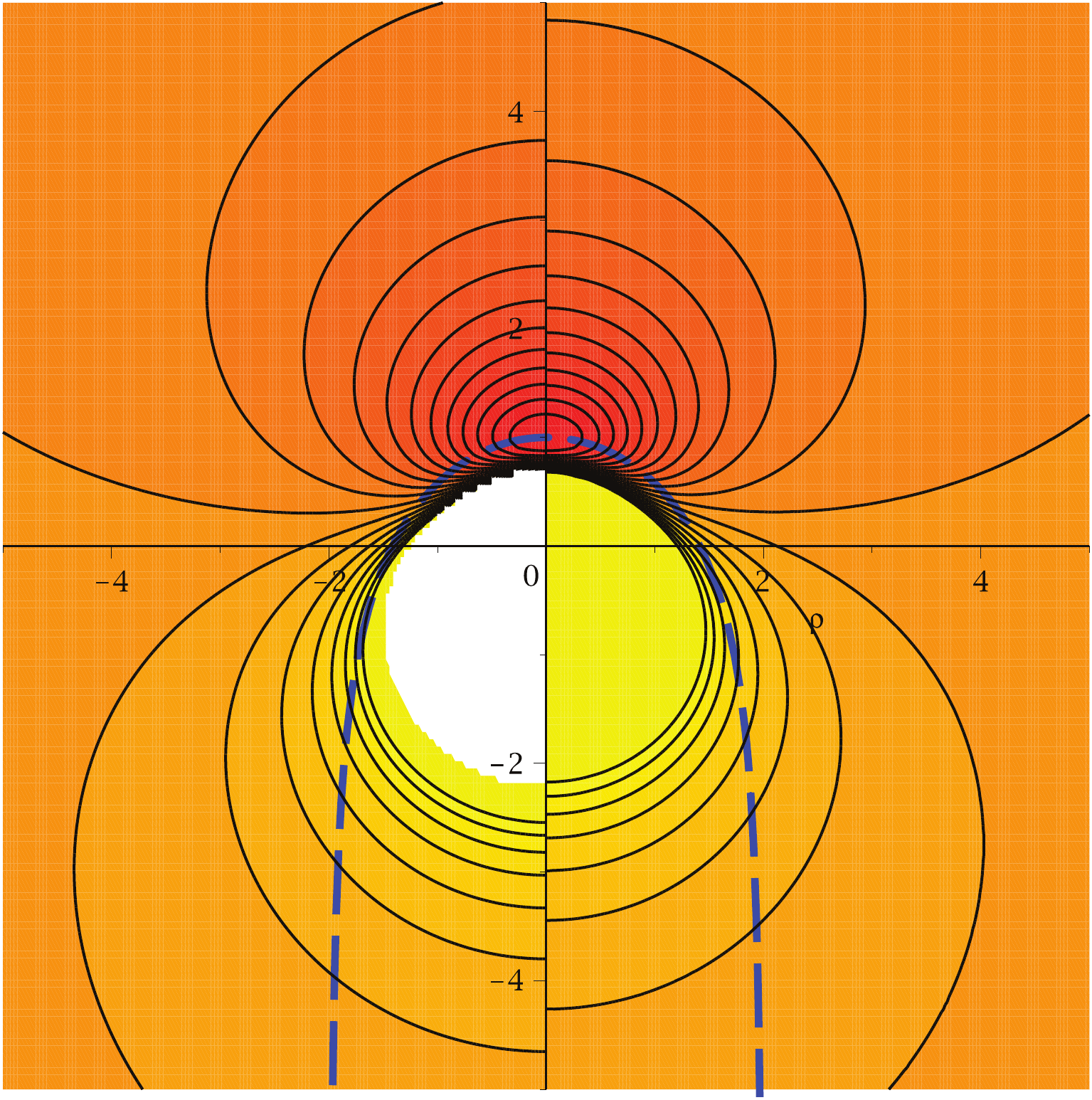}
  \caption{\label{fig:n-contours}
    Contours of density (for $\gamma=1$), comparing the exact solution
    (\ref{eq:imp_gamma_eq1}) on the left against its second-order
    approximation (\ref{eq:n_app_vals}) on the right side, shown for an
    upstream Mach number of $m=0.6$.  The dashed blue line marks the
    heliopause according to Eq.~(\ref{eq:def_hp}), and axis tick marks are
    in units of the stand-off distance $\sqrt{q}$.
    Contour values are spaced uniformly between $\nb_{\rm max} \approx 1.197$
    (red) and $\nb_{\rm min} \approx 0.875$ (yellow), and smaller values are
    capped.  The qualitative, and even quantitative, agreement is
    evidently very reasonable. \\
  }
\end{figure}
As can prominently be seen from the right plot, the approximation is not
at all restricted to subsonic flow, but does indeed feature a smooth sonic
transition, leading to moderately high local Mach numbers of up to about 1.83
before eventually slowing down again toward its original, subsonic speed.

Fig.~\ref{fig:n-contours} shows the density contours in $(\rho,z)$
space that result from the exact solution and its second-order approximation,
confirming the expectation of mass piling up in front of the heliopause and
being washed down the flanks.  The chosen value of $m=0.6$ is close to
$m_{\rm crit, isoth}$, beyond which the exact solution is no longer
well-defined everywhere outside the heliopause. However, we remark that Mach
numbers up to $m=1$ may be used for the approximation in the entire space and
the exact solution restricted to the upwind half-space near the inflow axis.
As expected, the density is highest at the stagnation point. Inside the white,
approximately circular region around the origin, no solution is available.
This region grows with larger $m$, eventually engulfing not only the
heliopause but the entire downwind half-space $z \le 0$ as $m \rightarrow 1$.
Since the approximative solution is a simple second-order polynomial, it
can be defined in the entire space for any $m$, and thus exhibits no such
white region.

It may at first sight seem questionable to rely on an ``approximation'' in a
region where the exact solution is not at all available. What happens is that
the approximation is first derived from the exact solution where it exists,
and subsequently continued into the ``white'' region in which the exact
solution is not defined, exploiting the fact that the former, due to its
simple polynomial form, is void of any domain restrictions. As a consequence,
we have no measure of the flow's departure from the exact momentum-conserving
solution in the continuation region. What we do know, however, is that it
proceeds along the same Rankine-type stream lines, conserves mass exactly,
and connects smoothly into the original solution's domain. Given that
momentum is not exactly conserved in either region anyway, we feel that
these characteristics are nevertheless sufficient for the flow field to
clearly qualify as physically meaningful for its intended purpose of deriving
an improved formula for the heliospheric interstellar magnetic field. In the
same vein, we note that none of the existing models
\citep[][etc.]{Whang:2010, Schwadron_EA:2014, Isenberg_EA:2015}
come anywhere near this degree of physical realism, despite their
unimpeachably acknowledged usefulness for the heliophysics community. In the
following, we will continue to refer to the thus extended approximation
simply as ``approximation,'' irrespectively of the position at which it is
evaluated.

\begin{figure}
  \includegraphics[width=8.5cm]{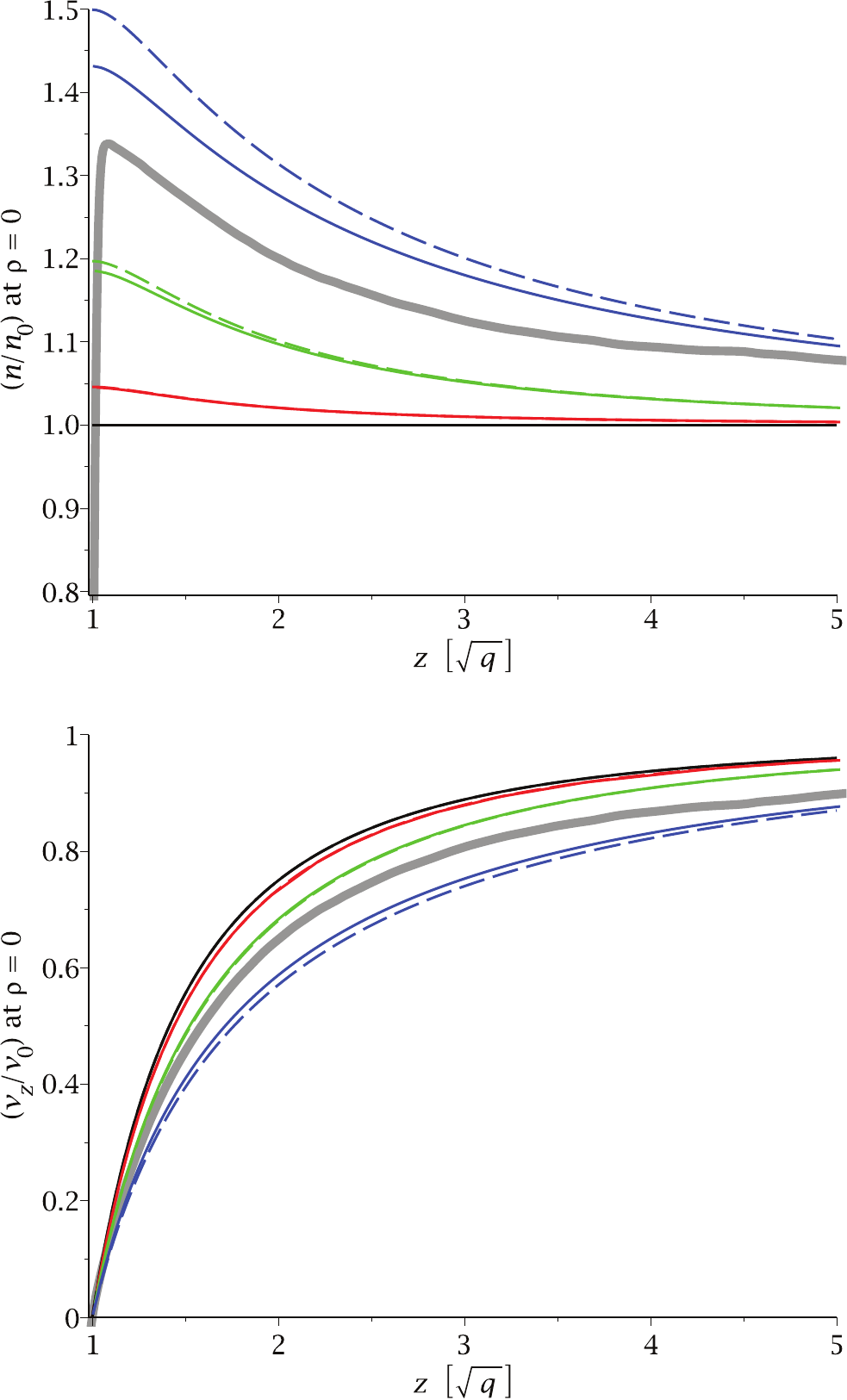}
  \caption{\label{fig:inflow_dens+velo}
    Comparison of normalized density (upper panel) and velocity (lower panel)
    along the inflow axis $(\rho=0)$ for upstream Mach numbers $m=0$ (black),
    $m=0.3$ (red), $m=0.6$ (green), and $m=0.9$ (blue) for both $\gamma=1$
    (solid) and $\gamma=5/3$ (dashed) as a function of normalized upstream
    heliocentric distance $z$, in units of $\sqrt{q}$. The thick, gray curves
    are extracted from a numerical simulation using $m=0.8$. For this case,
    the stand-off distance to the stagnation point, identified as the $z$
    distance at which $u_z$ vanishes, is found to equal
    \mbox{$\sqrt{q}=163$~AU}. The steep decline of $n/n_0$ toward the
    stagnation point is an artefact of finite numerical resolution, which
    causes the density outside the heliopause to approach the (much lower)
    corresponding value on the inside.
  }
\end{figure}

The inflow axis ($\rho=0$) is of particular interest, especially since the
full momentum equation is satisfied there. Therefore,
Fig.~\ref{fig:inflow_dens+velo} depicts the density and flow velocity along
this axis as functions of the upstream heliocentric distance $z$ for both the
isothermal ($\gamma=1$) and the adiabatic ($\gamma=5/3$) cases. The difference
between them is obviously negligible except for large $m$. A pile-up of mass
is apparent, which is of course absent from the incompressible case ($m=0$).
The flow also decelerates much stronger at high Mach numbers. Additionally
shown are the corresponding profiles obtained from a fully self-consistent
hydrodynamical simulation, details of which are described in
Section~\ref{sec:numerics_hd}. The normalized numerical peak density of
1.338 thus obtained is in excellent agreement with its theoretical value
of 1.337. Moreover, the value of $\bar{n}=1.078$ at $z=5 \sqrt{q}$ is only
slightly above 1.056, which is the normalized density predicted by
Eq.~(\ref{eq:imp_gamma_ne1}). \\

\section{Derivation of the Local Interstellar
  Magnetic Field}
\label{sec:mag-field}

\subsection{Cylindrical Field Components}

The components of the advected magnetic field ${\bf B}$ are derived using
the method of line conservation \citep[see][]{Elsasser:1956,
  Naor_Keshet:2015}, which applies the idea behind Cauchy's integral
(first formulated by \citet{Cauchy:1816} in the context of fluid mechanics)
to ideal MHD. We thus exploit the fact that in ideal MHD, the quantity
${\bf B}/n$ satisfies the same equation of motion as the line element
${\bf \delta x}$ connecting two neighboring particles that are passively
advected in the flow field ${\bf u}$.

While at upstream infinity,
$\left({\bf B}/n\right)|_{\infty} = {\bf B}_0/n_0$ is spanned by the
basis vectors $\{ {\bf e}_{\rho}, {\bf e}_{\varphi}, {\bf e}_z \}$, the
corresponding basis at finite position is
$\{ {\bf c}, (\rho/a) \, {\bf e}_{\varphi}, - \bar{\bf u} \}$, where ${\bf c}$
is a vector connecting two fluid elements that start at the same
``height'' $z_0$ on adjacent stream lines labelled by $a$ and $a+\delta a$,
and $a$ is a known function of position (cf.\ Fig.~2 and Eq.~(49) in Paper~I).
Since the basis vectors are co-moving with the flow, the coefficients  of
${\bf B}/n$ with respect to both bases are the same. We are therefore led to
\begin{equation}
  {\bf B}/n = \left[ B_{\rho 0} \, {\bf c}
    + B_{\varphi 0} (\rho/a) \, {\bf e}_{\varphi} - B_{z 0} \, {\bf \bar{u}}
  \right] / n_0 \ ,
\end{equation}
and thereby arrive at the relation
\begin{equation}
  \label{eq:bsol_comp}
  \begin{split}
    {\bf B}(\rho,\varphi,z) =&\ 
    \left[ B_{\rho 0} \ c_{\rho} - B_{z 0} \ \bar{u}_{\rho} \right]
    \bar{n} \ {\bf e}_{\rho} \\
    &+ B_{\varphi 0} (\rho/a) \, \bar{n} \ {\bf e}_{\varphi} \\
    &+ \left[ B_{\rho 0} \ c_z - B_{z 0} \ \bar{u}_z \right] \bar{n} \
    {\bf e}_z
  \end{split}
\end{equation}
that generalizes Eq.~(67) of Paper~I to the compressible case.
Furthermore, it was shown in Paper~I that the condition of equal travel times
\begin{equation}
  \label{eq:same-tt}
  \Int_{\rho_a}^{\rho} \frac{\dd \rho\p}{u_{\rho}(a,\rho\p)} = \Delta t =
  \Int_{\rho_{a+\delta a}}^{\rho+\delta \rho}
  \frac{\dd \rho\p}{u_{\rho}(a+\delta a,\rho\p)}
\end{equation}
can be used to derive the components of ${\bf c}$ as
\begin{eqnarray}
  \label{eq:def_crho}
  c_{\rho} &=& \frac{\delta \rho}{\delta a}
  = - \bar{u}_{\rho}(a,\rho) \ \frac{\partial}{\partial a} \int
  \frac{ \dd \rho }{\bar{u}_{\rho}(a,\rho)} \\
  \label{eq:def_cz}
  c_z &=& \frac{\delta z}{\delta a}
  = \frac{\partial z_a(\rho)}{\partial a} +
  \frac{\bar{u}_z}{\bar{u}_{\rho}} \, c_{\rho} \ .
\end{eqnarray}
The function $z_a(\rho)$ in Eq.~(\ref{eq:def_cz}) designates the $z$
coordinate of the flow line labelled by $a$ and parameterized by $\rho$,
which passes through $(\rho_a, z_a(\rho_a))$. Its derivative evaluates to
\begin{equation}
  \label{eq:dza_da}
  \frac{\partial z_a(\rho)}{\partial a} = \frac{a \, r^3}{q \, \rho^2}
\end{equation}
according to Eq.~(B.8) in Paper~I.

We now relate the magnetic field of the incompressible case (as derived in
and known from Paper~I) to the compressible case (with $m>0$). For this
purpose, we label the respective quantities with superscripts ``I'' and
``C,'' such that
\begin{eqnarray}
  \label{eq:Brho_CfromI}
  B_{\rho}^{\rm I,C} &=& - \left( B_{\rho 0} \, K_0^{\rm I,C} + B_{z 0} \right)
  \nb^{\rm I,C} \, \bar{u}_{\rho}^{\rm I,C} \\
  \label{eq:Bz_CfromI}
  B_z^{\rm I,C} &=& - \left( B_{\rho 0} \, K_0^{\rm I,C} + B_{z 0} \right)
  \nb^{\rm I,C} \, \bar{u}_z^{\rm I,C} \\
  && \nonumber + B_{\rho 0} \, \frac{\partial z_a}{\partial a} \,
  \nb^{\rm I,C} \ ,
\end{eqnarray}
where $\nb^{\rm C}=\nb$, $\nb^{\rm I}=1$, and
\begin{equation}
  \label{eq:def_K0}
  K_0^{\rm I,C} := \frac{\partial}{\partial a}
  \int \frac{\dd \rho}{\bar{u}_{\rho}^{\rm I,C}(a,\rho)} \ .
\end{equation}
In both cases, the momentum density is prescribed through the same potential
as ${\bf s}^{\rm I,C} =-\nabla \Phi$, implying
\begin{equation}
  n_0 \, {\bf u}^{\rm I} = {\bf s}^{\rm I,C} = n \, {\bf u}^{\rm C} 
  \quad \Leftrightarrow \quad
  \bar{\bf u}^{\rm C} = \nb^{-1} \, \bar{\bf u}^{\rm I} \ .
\end{equation}

Since $\nb$ (and thus $\bar{\bf u}^{\rm C}$) is given only implicitly, we may
not hope to obtain analytical expressions for the integral $K_0^{\rm C}$ in
Eqs.~(\ref{eq:Brho_CfromI}) and (\ref{eq:Bz_CfromI}). However, when
approximating the normalized density as
\begin{equation}
  \label{eq:n_approx_nu}
  \bar{n} = \nb_{\rm sp} + \nu_1 \A + \nu_2 \A^2
\end{equation}
with the values for $\nu_{1,2}$ read off from Eq.~(\ref{eq:n2_app}),
$K_0^{\rm C}$ can be related to $K_0^{\rm I}$ via
\begin{eqnarray}
  \label{eq:HIvsHC}
  K_0^{\rm C}(a,\rho) &=& \frac{\partial}{\partial a}
  \int \frac{\nb}{\bar{u}_{\rho}^{\rm I}(a,\rho)} \, \dd \rho \\
  \nonumber &=& \nb_{\rm sp} \, K_0^{\rm I}(a,\rho)
  + \underbrace{\nu_1 K_1^{\rm I}(a,\rho) +
    \nu_2 K_2^{\rm I}(a,\rho)}_{=:{\cal C}} \ ,
\end{eqnarray}
in which the definition
\begin{equation}
  \label{eq:def_K}
  K_k^{\rm I}(a,\rho) := \frac{\partial}{\partial a}
  \int \frac{\A^k}{\bar{u}_{\rho}^ {\rm I}(a,\rho)} \, \dd \rho
  \ , \ k \in \{1,2 \} \ ,
\end{equation}
generalizes that in (\ref{eq:def_K0}). This allows us to express the desired
magnetic field solution for the compressible case in terms of the known one
for the incompressible case as
\begin{eqnarray}
  \label{eq:BC_comps_rho}
  B_{\rho}^{\rm C} &=& \nb_{\rm sp}
  B_{\rho}^{\rm I} + \bar{u}_{\rho}^{\rm I} {\cal Z} \\
  B_{\varphi}^{\rm C} &=&  \nb \, B_{\varphi}^{\rm I} \\
  \label{eq:BC_comps_z} B_z^{\rm C} &=& \nb_{\rm sp}
  B_z^{\rm I} + \bar{u}_z^{\rm I} {\cal Z} + (\nb-\nb_{\rm sp}) B_{\rho 0} \,
  \frac{\partial z_a}{\partial a} \ ,
\end{eqnarray}
where
\begin{equation}
  \label{eq:def_calZ}
  {\cal Z} := (\nb_{\rm sp}-1) B_{z 0} - {\cal C} \, B_{\rho 0} \ .
\end{equation}
The incompressible case is evidently recovered in the limit
$m \rightarrow 0$, as in that case
$\nb_{\rm sp} \rightarrow 1$, $\nb \rightarrow 1$, ${\cal C} \rightarrow 0$,
and ${\cal Z} \rightarrow 0$.

We now determine an explicit expression for the correction term ${\cal C}$.
In order to evaluate the integrals $K_{1,2}$, we first consider the normalized
flow potential \mbox{$\bar{\Phi}=\bar{\Phi}(\rho,z_a(\rho))$} along the
streamline in question. The total $\rho$ derivative of this quantity becomes
\begin{equation}
  \label{eq:dPhi_drho}
  \begin{split}
    \frac{\dd \bar{\Phi}}{\dd \rho}
    =&\ \frac{\partial \bar{\Phi}}{\partial \rho}
    +   \frac{\partial \bar{\Phi}}{\partial z} \frac{\dd z}{\dd \rho}
    =   \frac{\partial \bar{\Phi}}{\partial \rho} +
    \left( \frac{\partial \bar{\Phi}}{\partial z} \right)^2
    \left( \frac{\partial \bar{\Phi}}{\partial \rho} \right)^{-1}
    \\
    =& \underbrace{\left[
        \left( \frac{\partial \bar{\Phi}}{\partial \rho} \right)^2 +
        \left( \frac{\partial \bar{\Phi}}{\partial z}    \right)^2
      \right]}_{=(\nabla \bar{\Phi})^2 = \, \bar{\bf s}^2}
    \left( \frac{\partial  \bar{\Phi}}{\partial \rho} \right)^{-1}
    = \frac{\A}{- \bar{u}_{\rho}^ {\rm I}} \ .
  \end{split}
\end{equation}
For $k=1$, we thus simply obtain
\begin{equation}
  \label{eq:def_K1}
  K_1^{\rm I}(a,\rho) = - \frac{\partial}{\partial a}
  \int \frac{\dd \bar{\Phi}}{\dd \rho} \, \dd \rho
  = - \frac{\partial \bar{\Phi}}{\partial a}
  = \frac{(r^3-q z) \, a}{q \, \rho^2} \ ,
\end{equation}
where
\begin{eqnarray}
  \label{eq:barPhi}
  \bar{\Phi} &=& \frac{q}{\sqrt{\rho^2+z_a(\rho)^2}} + z_a(\rho) \\
  &=& \frac{2 \rho^2-a^2}{2\rho} \sqrt{ \frac{4 q + a^2-\rho^2}{\rho^2-a^2}}
  -\frac{\rho}{2} \sqrt{\frac{\rho^2-a^2}{4 q + a^2-\rho^2}} \ . \nonumber
\end{eqnarray}
The integral for $k=2$ requires explicit evaluation. Using
Eqs.~(\ref{eq:dPhi_drho}) and (\ref{eq:barPhi}) yields, after lengthy but
straightforward computation,
\begin{equation}
  \label{eq:def_K2}
  \begin{split}
    K_2^{\rm I}(a,\rho) &= \frac{\partial}{\partial a} \int \bar{u}_{\rho}
    \left(\frac{\dd \bar{\Phi}}{\dd \rho} \right)^2 \dd \rho \\
    &= \frac{\partial}{\partial a}
    \left( - \frac{q^3}{5 \, r^5} +  \frac{q^2 z}{r^4} - \frac{3q}{r} - z
    \right) \\
    &= \frac{a}{\rho^2} \left( 
      \frac{q^2 z}{r ^4} + \frac{(\rho^2-3z^2)q}{r^3} + 3 z
      - \frac{r^3}{q} \right) \ .
  \end{split}
\end{equation}

At this point, all prerequisites needed for the explicit construction of the
'compressible' magnetic field components $B^{\rm C}_{\rho, \varphi, z}$ from the
known 'incompressible' ones $B^{\rm I}_{\rho, \varphi, z}$ are in place, and
this construction proceeds as follows.
\begin{enumerate}
\item Choose an upstream Mach number $m < 1$ (preferably near unity).
  The fast magnetosonic Mach number might serve as an educated guess
  \citep{Spreiter_Stahara:1995}.
\item Compute the constant coefficients $\nb_{\rm sp}$ and $\nu_{1,2}$
  using Eqs.~(\ref{eq:n_max}) and either (\ref{eq:n1_app}) or
  (\ref{eq:n2_app}).
\item At each desired position $(\rho, \varphi, z)$, evaluate
  \begin{itemize}
  \item[(a)] functions $K_{1,2}$ from Eqs.~(\ref{eq:def_K1}) and
    (\ref{eq:def_K2}),
  \item[(b)] functions ${\cal C}$ and ${\cal Z}$ from Eqs.~(\ref{eq:HIvsHC})
    and (\ref{eq:def_calZ}),
  \item[(c)] function $\partial_a z_a(\rho)$ from Eq.~(\ref{eq:dza_da}).
  \end{itemize}
\item Finally, substitute all quantities into
  Eqs.~(\ref{eq:BC_comps_rho})--(\ref{eq:BC_comps_z}) to arrive at the
  desired field components, bearing in mind that
  \mbox{${\bf u}^{\rm I} = -\nabla \Phi$} according to Eq.~(\ref{eq:u-flow}).
\end{enumerate}

In principle, the approximation (\ref{eq:n_approx_nu}) could be continued to
even higher orders in $\A$. However, tentative computations indicate that the
expressions for $K_k^{\rm I}$ become much more involved for $k>2$, and that
comparatively little could be gained by going to order $\A^3$ or higher.
We concede that the presented procedure is already more involved than the
original field derived in Paper~I, but also note that the result is clearly
more realistic, not only because it satisfies more physical constraints, but
also because it does indeed perform better in direct comparison to a
numerical model.
This will be demonstrated in Section~\ref{sec:numerics}. \\

\subsection{Cartesian Field Components on the Inflow Axis}

The cylindrical components (\ref{eq:BC_comps_rho})--(\ref{eq:BC_comps_z})
of the magnetic field can easily be converted into a Cartesian representation.
The only region for which this conversion is not straightforward is the
$z$~axis, where both $B_{\rho,\varphi}$ and the transformation factors
$\sin\varphi$ and $\cos\varphi$ are ill-defined, and the limit
$\rho \rightarrow 0$ is somewhat cumbersome to evaluate. But since the
conditions along the inflow axis are of particular physical interest
(especially given that they satisfy the momentum conservation equation
(\ref{eq:momentum}) exactly), it is fortunate that an easier avenue is
available to treat this special case. For this purpose, we revert to the
induction equation (\ref{eq:induct}), now written in the form
\begin{equation}
  \nabla \times \big[ ({\bf s}/n) \times {\bf B} \big] = {\bf 0} \ ,
\end{equation}
which is equivalent to
\begin{equation}
  \label{eq:induct_compress}
  \big[ ( {\bf s} \cdot \nabla n) {\bf B}
  - ( {\bf B} \cdot \nabla n) {\bf s} \big] /n =
  ({\bf s} \cdot \nabla) {\bf B} - ({\bf B} \cdot \nabla) {\bf s}
\end{equation}
due to $\nabla \cdot {\bf s} = 0 = \nabla \cdot {\bf B}$. On the $z$ axis,
where the only non-zero components of both ${\bf s}$ and $\nabla n$ are those
in $z$ direction, the left-hand side of Eq.~(\ref{eq:induct_compress})
simplifies to
\begin{equation}
  (B_x {\bf e}_x + B_y {\bf e}_y) \, s_z\, \partial_z(\ln n) \ .
\end{equation}
The Cartesian components of the right-hand side evaluate to
\begin{equation}
  s_0 \left( \frac{q}{z^2}-1 \right)
  \left( \begin{array}{c}
      \partial_z B_x \\  \partial_z B_y \\  \partial_z B_z \\
    \end{array} \right) + \frac{s_0 \, q}{z^3}
  \left( \begin{array}{c} -B_x \\ -B_y \\ 2 \, B_z \\ \end{array} \right) \ .
\end{equation}
In total, we obtain the differential equations
\begin{eqnarray}
  \partial_z \ln \left(\frac{B_x}{n}\right) &=& \frac{q}{z \, (q-z^2)} =
  \partial_z \ln \left(\frac{B_y}{n}\right) \\
  \partial_z \ln \left(      B_z    \right) &=& \frac{-2 q}{z \, (q-z^2)} \ ,
\end{eqnarray}
which are readily solved to yield the remarkably simple expressions
\begin{eqnarray}
  \label{eq:Bxy_on_axis}
  \frac{B_x|_{\rho=0}}{B_{x 0}} &=& \nb|_{\rho=0}
  \left(1-\frac{q}{z^2}\right)^{-1/2} = \frac{B_y|_{\rho=0}}{B_{y 0}} \\
  \label{eq:Bz_on_axis}
  \frac{B_z|_{\rho=0}}{B_{z 0}} &=& 1-\frac{q}{z^2}
\end{eqnarray}
as a generalization of the corresponding formulas from Appendix~C of
Paper~I.

It is interesting to note that, while $B_x$ and $B_y$ each contribute a
factor $\nb$ in the compressible case, $B_z$ is identical to its
incompressible analog. This observation may at first sight come as a
surprise, but is indeed consistent with the notion of line conservation:
For the distance $\delta z$ between two neighboring particles both travelling
along the inflow axis, we have
\mbox{$B_z/n \propto \delta z \propto u_z$} and hence
\mbox{$B_z \propto n \, u_z = s_z = -\partial_z \Phi$}, which is the same in
both cases. On the other hand, the separation $\delta x$ between two
particles travelling on adjacent flow lines at the same height $z$ only
depends on the geometry of the flow lines at that height, which is determined
by $s_z$, hence \mbox{$B_x/n \propto \delta x \propto s_x$}, so
\mbox{$B_x \propto n \, s_x$}. \\

\section{Comparison to (M)HD Simulations}
\label{sec:numerics}

Whenever the need for a prescription of an interstellar magnetic field arises
in the heliospheric context, it is obtained either analytically or by way of
numerical simulations. Therefore, it seems reasonable to asses the
usefulness of our new model by comparing it to a field that would typically
be generated using a self-consistent MHD code. In the same vein, it is 
of interest to investigate to what extent the hydrodynamical aspects of this
model are compatible with corresponding simulation data.

\subsection{Pure HD for Flow Structure}
\label{sec:numerics_hd}

We first provide details of the simulation underlying the gray plot profiles
in the panel of Fig.~\ref{fig:inflow_dens+velo}. For these, the grid-based
MHD code {\sc Cronos} was employed to simulate an unmagnetized, axisymmetric
heliosphere. The poloidal $(\rho,z)$ domain of size
$[0,2000]$~AU$\, \times\, [-1000,1000]$~AU was covered with cells of size
$\Delta\rho=1$~AU for $\rho < 1000$~AU and 5~AU for $\rho \in [1000,2000]$~AU,
and $\Delta z=1$~AU throughout. The outer layer of coarser cells served to
shield the inner part from reflections possibly emanating from this boundary.
Parameters were those from the plasma-only version of the
\citet{Mueller-EA:2008} benchmark (cf.\ Table~1 in that paper), except that
the ISM temperature was raised from 6530~K to 39553~K in order to achieve the
desired upstream Mach number of $m=0.8$. Such high temperature is consistent
with numerical findings \citep[see, e.g.,][]{Pauls_Zank:1997, Fahr_EA:2000}
that take explicitly into account a shock transition and charge exchange
processes. The displayed profiles were extracted at simulation time $t=1500$
(in units of 1~AU$/c_{\rm s}$, corresponding to about 280~years in physical
units), at which the configuration was deemed sufficiently stationary by
visual inspection. \\

\subsection{Full MHD for Magnetic Structure}

The simulation that we employ to assess the quality of our improved magnetic
field solution, as well as the resulting data, are identical to those used in
Paper~I. In particular, the undisturbed interstellar magnetic field of
strength 0.3~nT is oriented in the \mbox{$x$--$z$} plane and makes an angle
of $50^{\circ}$ with the inflow axis. For further details of the numerical
setup, the reader is again referred to that paper.

\begin{center}
  \begin{figure*}[h!]
    \includegraphics[width=\textwidth]{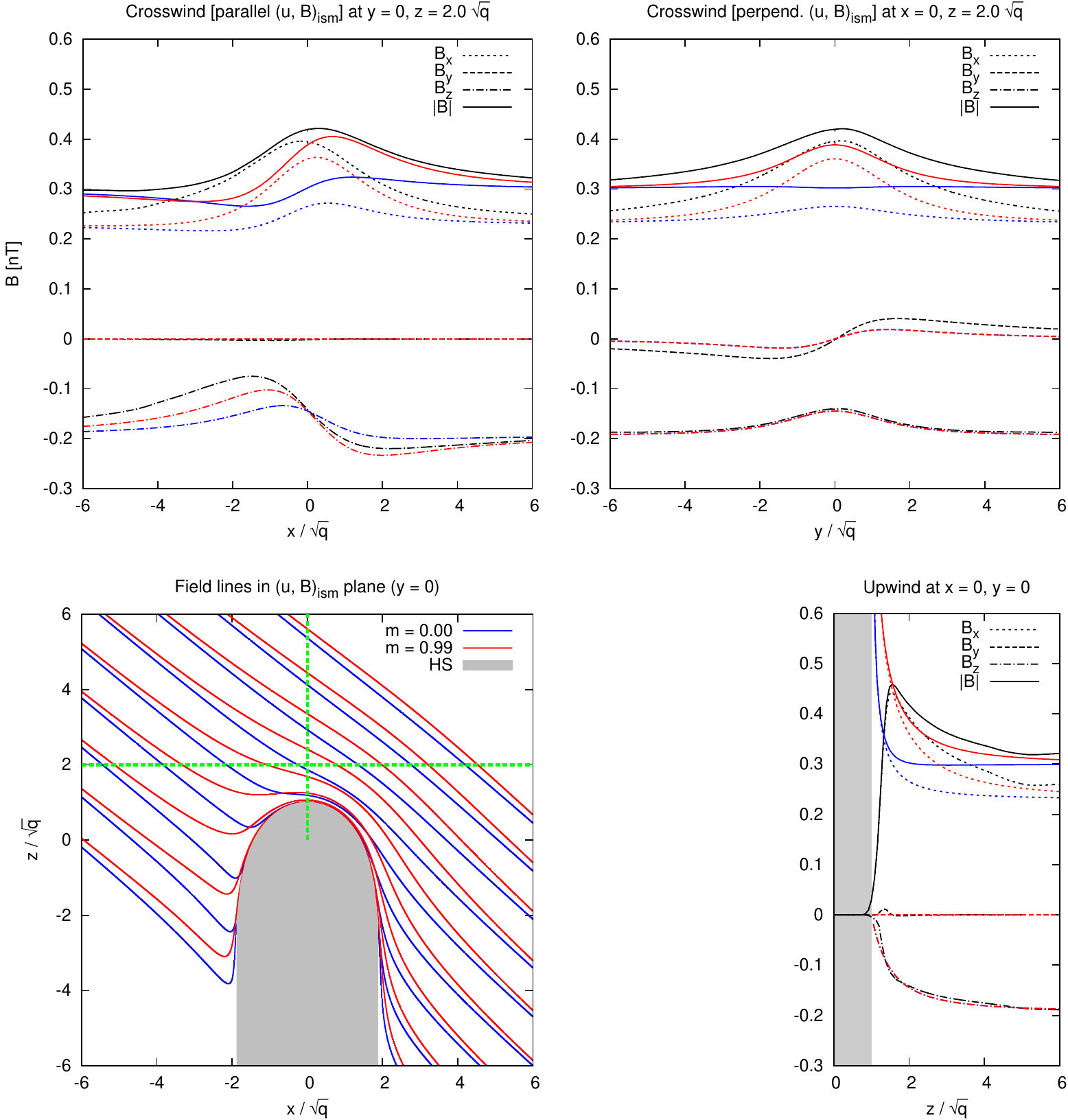}
    \caption{ \label{fig:comp_panel}
      Cartesian magnetic field components $B_x$ (dotted), $B_y$ (dashed),
      $B_z$ (dashed-dotted), and $\|{\bf B}\|$ (solid) along lines passing
      through the point $(x,y,z)=( 0,0,2) \sqrt{q}$, comparing the
      incompressible ($m=0$, blue) to the (almost maximally) compressible
      ($m=0.99$, red) case. Additionally, the corresponding field components
      that were extracted from the numerical simulation are shown in black.
      Note that in the upper right plot, the $B_y$ and $B_z$ curves for
      $m=0.99$ only seem to be missing because they coincide with those for
      $m=0$, as can be seen from Eqs.~(\ref{eq:Bxy_on_axis}) and
      (\ref{eq:Bz_on_axis}).
      In the lower left quadrant, selected field lines for both models can
      be seen draping around the heliosphere (viewed along the positive $y$
      axis). Additionally, the green dashed lines in this plot indicate the
      position at which the cuts along $x$ and $z$ are taken.
    }
  \end{figure*}
\end{center}

Fig.~\ref{fig:comp_panel} compares all magnetic field components of the
numerical solution to its isothermal, second-order, analytical counterpart
along straight lines parallel to the Cartesian $x$, $y$, and $z$ axes and
passing close to the stagnation point.
The superior performance of our improved compressible model over its
incompressible predecessor is evident from the observation that in all cases,
the red curves yield much better approximations to the numerical data (black)
than the blue ones. Together with the field line plot in the lower left
quadrant, this panel illustrates and confirms the expected differences
between the incompressible and the compressible model:
In the compressible case,
the heliosphere's region of influence extends much further out in the
upstream direction, the field lines are less sharply curved, and the field
strength amplification resulting from magnetic pile-up ahead of the
stagnation point is more pronounced. All of these characteristics contribute
to a markedly improved degree of realism and physical usefulness. \\
\vspace*{0mm}

\section{Summary and Conclusions}
\label{sec:summary}

In this work, the exact analytical MHD model by \citet{Roeken_EA:2015} for
the interstellar magnetic field in the outer vicinity of the heliopause has
been significantly extended from incompressible to compressible, yet
predominantly subsonic flow, with the upstream sonic Mach number $m < 1$ as a
free parameter. Although the use of flow potentials precludes the occurrence
of a bow shock, the model's qualitative applicability may well be extended
to supersonic but sub-alfv\'enic flows, since it should be of minor relevance
whether the field lines are distorted due to thermal or magnetic pressure.

This improved model transcends its predecessor not only conceptually by
approximately satisfying conservation of linear momentum along stream lines,
but also yields results that are much closer to those from a self-consistent,
grid-based MHD simulation.

While the derived spatial structures of flow lines and density profiles are
exact for both isothermal and adiabatic settings, the corresponding solution
for a magnetic field being passively frozen into this flow required the exact
density solution to be replaced by a suitable polynomial approximation to
warrant analytical tractability.
As an additional benefit, this approximation allows itself to be continued
into spatial regions in which the exact solution is not defined. And unlike
the exact flow solution, its continued approximation is not restricted to
subsonic flows, but features a smooth sonic transition
\citep[reminiscent of ``sonic lines,'' e.g.,][]{Scherer_EA:2016} in the
heliotail's flanks.

Furthermore, for the velocity field associated with this approximative
density distribution, the derived magnetic field is again an exact, solenoidal
solution to the induction equation of ideal MHD. In particular, the magnetic
field's orientation at the upstream boundary can still be chosen freely.
But even disregarding the entire magnetic part of the model, its
hydrodynamic configuration can still offer a marked improvement over models
in which incompressible or even compressible but isothermal flows are
prescribed \citep{Parker:1961, Nerney_Suess:1995}.
We note in passing that, although this option was not exploited in the
present work, a physically meaningful temperature distribution of the local,
heliosphere-dominated interstellar medium becomes readily accessible from
the ideal gas law and the density field, which our model can provide.

\acknowledgments

We acknowledge financial support via the projects FI~706/15-1 and
FI~706/21-1 funded by the
Deutsche Forschungsgemeinschaft (DFG), as well as through the
{\em RAPP (Ruhr Astroparticle and Plasma Physics) Center}, funded
as MERCUR project St-2014-040.  Furthermore, we also thank the anonymous
referee for helpful and constructive comments.

\bibliographystyle{apj}
\bibliography{references_compress}

\end{document}